\documentclass[10pt,showpacs]{revtex4}
\usepackage{amssymb,epsf}
\usepackage{latexsym}

\begin{document}

\title{Dynamic conformal spherically symmetric solutions in an accelerated background}
\author{H. Moradpour$^1$\footnote{h.moradpour@riaam.ac.ir} and N. Riazi$^2$\footnote{n$\_$riazi@sbu.ac.ir}}
\address{$^1$ Research Institute for Astronomy and Astrophysics of Maragha (RIAAM),
P.O. Box 55134-441, Maragha, Iran,\\
$^2$ Physics Department, Shahid Beheshti University, Evin, Tehran 19839, Iran.}

\begin{abstract}
We consider dynamical spherically symmetric spacetimes, which are
conformal to the static spherically symmetric metrics, and find new
solutions of Einstein equations by symmetry considerations. Our
study help us classify various conformal Black Holes that are
embedded within a dynamic background into the one class of solutions
with the same conformal symmetry. In addition, Thermodynamics,
mathematical and gravitational properties are addressed. These
solutions point to have a better resolution of the meaning of the
Black Holes in the dynamic background.
\end{abstract}

%\keywords{Conformal transformation; perfect fluid; dynamical black
%holes; black hole thermodynamics.}

\pacs{04.20.-q; 04.70.-s.}

\maketitle

\section{Introduction\label{sec1}}
Exact solutions of the Einstein equations attract more
investigations because they allow a global perception \cite{b1}.
One important issue in the relativistic astrophysics is the
evolution of the stars, which needs the interior solutions of
Einstein equations and help us find useful signals of the life of
the cosmos. Supernova explosions as one possible form of final
stage of stellar evolution is an example. Although a star is a
dynamical system, because of stellar tangible long time evolution,
one can use static solutions for studying the system so, the
spherically symmetric static metrics have a particular place in
the subject. Some of this significance is due to the symmetry
considerations in the stellar formation and evolution. One can
write general form of the static spherically symmetric metrics as:
\begin{eqnarray}\label{ssm}
ds^2=-e^{\nu}d\overline{t}^2+e^{\lambda}dr^2+e^{\mu}d\Omega^2,
\end{eqnarray}
where $\nu$, $\lambda$ and $\mu$ are functions of radius $(r)$,
only \cite{b2,sn02,b3,P}.

General static solution for isotropic fluid spheres, charged
perfect fluid version and certain types of dynamic metrics are
presented in the literature \cite{b4,b40,b401}. Diaz and Pullin
have found solutions for spheres with slow rotation \cite{DP}. In
the collapsing procedure, density increases and there are various
physical phenomena which induce anisotropy \cite{ra,ra0}, Rago has
generalized solutions to anisotropic static fluids \cite{b5}. The
radii at which $e^{\nu(r)}=0$, points to horizons \cite{P} which
obey some special laws. Nowadays these laws, which originally were
claimed for static Black Holes (BHs) and were originally proposed
by Bekenstein and Hawking \cite{BH,BH0,BH01}, are recognized as
backbone of thermodynamical properties of gravity
\cite{ja,ja0,cai,cai0,ver}. Exterior solutions of Einstein field
equations point us to the effects of the material content of the
universe on the background, which is now accepted as an
accelerating spacetime. Considering cosmological principle
\cite{Roos}, background spacetime can be expressed by conformal
form of the so called Friedmann-Rabertson-Walker (FRW) model
\cite{confrw}
\begin{eqnarray}\label{FRW}
ds^2=a(\eta)^2[d\eta^2+dr^2+r^2d\Omega^2],
\end{eqnarray}
where $a(\eta)$ is scale factor in the conformal time $(\eta)$
notation. The conformal form of the FRW metric enables one to
incorporate spatial inhomogeneity via the conformal factor
\cite{R}.

Considering the above arguments, it is now clear that finding a
general form of non static spherically symmetric metrics is
desirable and in fact it has attracted some interest. There are
four independent approaches for the task. In the first approach,
by focusing on symmetry considerations which help us to simplify
the Einstein equations, some authors try to find solutions to the
Einstein field for static and non-static fluids
\cite{HP,MMT,MMM,MSM,PPB}. More solutions including isotropic and
anisotropic fluids can be found in the references
\cite{more,more0,more01,more02,more03,more04,more05}.

In the second approach, in order to find the effects of the cosmic
expansion on the collapsing systems (specially, spherically
symmetric systems) work started by Einstein et al. Authors tried
to connect Schwarzschild solution to FRW on the boundary by
satisfying junction conditions and now, this solution is
classified as a more general model named Swiss Cheese model
\cite{ES,kan1,dro,saida,P1} attracted more investigations to
itself \cite{grenon,kan2,flory}. Finally, we should note that the
Swiss Cheese models can be classified as a subclass of
inhomogeneous Lema$\widehat{\textmd{i}}$tre-Tolman-Bondi models
\cite{SD',thes}.

In the third approach, some authors have embedded spherically
symmetric solutions into the FRW background and argued about their
surprising corollaries \cite{SD,SD0,SD01,SD'}. These solutions
include the Schwarzschild and Reissner-Nordst\"{o}m BHs in various
coordinate systems which lay into the FRW spacetime by a conformal
factor, which is compatible with the cosmic expansion eras. Since
these solutions are conformal to the Schwarzschild and
Reissner-Nordst\"{o}m spacetimes, their corresponding causal
structure remain the same. \cite{wald}. For these conformal
spacetimes, redshift singularities point to the expanding null
hypersurfaces which have non-zero confined surface area and cover
the BH curvature singularity ($r=0$) \cite{R'}. These
hypersurfaces change the causal structure of the metric the same
way as in the primary static metric. Among various conformal
spacetimes, the curvature scalars do not diverge at redshift
singularity only for Sultana and Dyer solution \cite{SD0}. In
addition, the energy conditions is problematic in their solution
\cite{SD'}. From what we have said and the fact that these objects
are the conformal transformation of static BHs, it is accepted
that some conformal models, such as Thakurta spacetime and
solutions by M$^{\textmd{c}}$Clure and Dyer, include dynamical BHs
\cite{SD'}. Also, only conformally Schwarzschild solution
(Thakurta spacetime) and solution by M$^{\textmd{c}}$Clure et al.
satisfy the energy conditions \cite{SD',R'}. In continue,
conformally Schwarzschild and Reissner-Nordst\"{o}m solutions can
be thought as a special group of metrics that have a Ricci scalar
that is conformal with the Ricci scalar of the FRW and may include
various horizons with different temperatures \cite{R'}.

In addition, some authors have tried to find dynamical BHs by
using the isotropic shape of the FRW metric along as perfect fluid
concept \cite{M1,G10}. For these solutions, mass and charge will
decrease as the functions of the universe expansion \cite{G10}.
Also, there is a hypersurface that acts like as an event horizon
which collapses while the background expands and its radius
depends on the curvature of the background. In addition, the
curvature scalars diverge on that. Also unlike the Swiss Cheese
models, energy conditions are violated by these solutions
\cite{SD'}. These features look unsatisfactory parts of the fourth
attempts. Therefore, these solutions do not contain dynamic BHs
\cite{SD',MG1,MG10,MG101,MG102}. Considering the prefect fluid
concept as well as the dynamic background~(\ref{FRW}), one can get
some solutions which include constant mass, charge and
cosmological constant \cite{mr}. In addition, the redshift
singularity is independent of the background curvature which is in
agreement with the FRW background, and points to the horizon-like
hypersurfaces \cite{mr}. More studies in which the prefect fluid
concept is used to derive dynamic spherically symmetric solutions
can be found in \cite{rosen1,rosen2}.

In this article, by following symmetry considerations, we want to
derive the various possible solutions of non static spherically
symmetric metrics. Throughout this paper we set the Einstein
gravitational constant ($k$) to one for simplicity ($k\equiv8 \pi
G$=1). The paper is organized as follows: in the next section, by
considering a general conformal killing vector and a non static
spherically symmetric metric, where metric dynamics comes from a
conformal factor which is only a function of time, we try to find
new solutions of the Einstein equations and we study their
physical and mathematical properties. Sections ($3$) includes
solutions with the BHs that merge into the dynamic background and
thermodynamics of these solutions. The last section is devoted to
a summary and concluding remarks.
%%%%%%%%%%%%%%%%%%%%%%%%%%%%%%%%%%%%%%%%%%%%%%%%%%%%%%%%%%%%%
\section{Conformal Spherically symmetric spacetimes\label{sec2}}
We begin by conformal form of (\ref{ssm}), where the conformal
factor has only time dependency:
\begin{eqnarray}\label{csm1}
ds^2=a(\eta)^2[-e^{\nu}d\eta^2+e^{\lambda}dr^2+e^{\mu}d\Omega^2],
\end{eqnarray}
where $\nu$, $\lambda$ and $\mu$ have only $r$ dependency,
$d\Omega^2=d\theta^2+\sin(\theta)^2 d\phi^2$ is the ordinary line
element on the unit $2$-sphere and $\eta$ is called conformal
time. We define cosmic time $t$ as usual:
\begin{eqnarray}\label{ct}
\eta \rightarrow t=\int a(\eta)d\eta.
\end{eqnarray}
Using the cosmic time coordinate, we obtain
\begin{eqnarray}\label{csm2}
ds^2=-e^{\nu}dt^2+a(t)^2[e^{\lambda}dr^2 +e^{\mu}d\Omega^2].
\end{eqnarray}
Since $n_{\alpha}=\delta^r_{\alpha}$ is the normal to the
hypersurface $r=const$, we have
\begin{eqnarray}\label{nhs}
n_{\alpha}n^{\alpha}=g^{rr}=\frac{e^{-\lambda}}{a(t)^2},
\end{eqnarray}
which is timelike when $e^{-\lambda}<0$, null for $e^{-\lambda}=0$
and spacelike if we have $e^{-\lambda}>0$. Therefore, it is
apparent that the existence of the null horizons is independent of
the functional time dependence of a non-zero scale factor
($a(t)$). Indeed, the causal structure of metric~(\ref{ssm}) is
invariant under the conformal transformation \cite{wald}. For a
co-moving observer, redshift of a radial incoming wave at the
point $(t,r)$ when it has been sent from $(r_0,t_0)$ is evaluated
as:
\begin{eqnarray}\label{rs}
1+z=\frac{\lambda(r,t)}{\lambda(r_0,t_0)}=\frac{a(t)}
{a(t_0)}(\frac{e^{\nu(r)}}{e^{\nu(r_0)}})^{\frac{1}{2}}.
\end{eqnarray}
It is seen that the redshift arises from two factors, one due the
cosmic expansion and one due to the local inhomogeneity. One also
obtains the following relation for the Ricci scalar:
\begin{eqnarray}\label{RS}
R=\frac{R_{FRW}}{e^{\nu}}+\frac{R_1}{2a(t)^2},
\end{eqnarray}
where, we have
\begin{eqnarray}\label{RFRW}
R_{FRW} = 6 \frac{a(t)\ddot{a}(t)+\dot{a}^2(t)}{a(t)^2},
\end{eqnarray}
and
\begin{eqnarray}\label{RS1}
R_1=e^{-\lambda}[2\mu'(\lambda'-\nu')+\nu'(\lambda'-\nu')
-3\mu'^2-2\nu''- 4\mu''] + 4e^{-\mu}.
\end{eqnarray}
When $(\dot{})$ and $(')$ are derivatives related to $t$ and $r$,
respectively. For $\nu=\lambda=0$ and $\mu=lnr^2$, we get $R_1=0$
and $R_s=R_{FRW}$. It is obvious that, solutions with $R_1=0$ have
Ricci scalar proportional to FRW's. By defining physical radius
$\zeta$ as
\begin{eqnarray}\label{ah1}
\zeta\equiv a(t)r
\end{eqnarray}
and introducing apparent horizon as a trapping surface with null
tangent from \cite{SW}, for the apparent horizon radius and its
surface gravity we get
\begin{eqnarray}\label{ah2}
\partial_{\alpha}\zeta\partial^{\alpha}\zeta=0\rightarrow r_H
\end{eqnarray}
and
\begin{equation}\label{SG}
\kappa=\frac{1}{2\sqrt{-h}}\partial_{a}(\sqrt{-h}h^{ab}\partial_{b}\zeta).
\end{equation}
Here $h_{ab}=diag(-e^{\nu},a^2(t) e^{\lambda})$ is induced metric
on the two dimensional hypersurface with $d\theta=d\phi=0$ and
temperature on this surface is $T=\frac{\kappa}{4\pi}$. For this
spherically symmetric spacetime, the confined Misner-Sharp mass
inside radius $\zeta$ is \cite{MS}:
\begin{eqnarray}\label{msm}
M=\frac{\zeta}{2}(1-h^{ab}\partial_a \zeta \partial_b \zeta).
\end{eqnarray}
When the apparent horizon is concerned $(h^{ab}\partial_a \zeta
\partial_b \zeta=0)$, this relation reduces to $M=\frac{\zeta}{2}$
which is equal to $M=\rho V$ for the FRW universe \cite{cai,cai0}.
The Einstein equations $(G_{\alpha\beta}=T_{\alpha\beta})$ leads
us to an anisotropic fluid $(P_r \neq P_T)$, which supports this
spacetime. We define the anisotropy function $\delta$ as:
\begin{eqnarray}\label{delta}
\delta\equiv
a(t)^2(P_r-P_T)=\frac{e^{-\lambda}}{4}[\nu'(\mu'+\lambda'-\nu')
+\lambda'\mu'-2\nu''-2\mu''-\frac{4e^{\lambda}}{e^{\mu}}],
\end{eqnarray}
and therefore, condition $\delta=0$ yields the isotropic
solutions. The only off diagonal elements of the Einstein tensor
are
\begin{eqnarray}\label{ODE}
G_{rt}=G_{tr}=H(t)\nu'.
\end{eqnarray}
In the above equation $H(t)\equiv\frac{\dot{a}(t)}{a(t)}$. As
previously mentioned, the FRW results can be reproduced by
choosing $\nu=\lambda=0$ and $\mu=2lnr$. For every observer with
four velocity $U^{\alpha}$, the stress-energy tensor can be
decomposed as \cite{EE}
\begin{eqnarray}
T_{\mu \nu}=\rho U_{\mu}U_{\nu} + Ph_{\mu \nu} + \Pi_{\mu \nu} +
q_{\mu}U_{\nu}+ q_{\nu}U_{\mu}.
\end{eqnarray}
In this equation, $\rho=T_{\mu \nu}U^{\mu}U^{\nu}$ plays the role
of the energy density and $h_{\mu \nu}\equiv g_{\mu
\nu}+U_{\mu}U_{\nu}$ is a projection tensor. In addition,
$P=\frac{1}{3}h_{\mu \nu}U^{\mu}U^{\nu}$ and $\Pi_{\mu
\nu}=T_{\alpha \beta} h^{\alpha}_{(\mu}h^{\beta}_{\nu)}$ are the
isotropic pressure and the traceless stress tensor respectively.
$q^{\mu}$ is the energy flux (the momentum density) relative to
$U^{\mu}$ when $q^{\mu}<0$ ($q^{\mu}>0$) signifies the input
(output) flow \cite{EE,Gao2}. For a co-moving observer
($U^\mu=|-g^{00}|^{1/2}\delta^\mu_t$), using the Einstein
equations, We get
\begin{eqnarray}\label{EMT2}
T_{\mu \nu}=\rho U_{\mu}U_{\nu} + Ph_{\mu \nu} + q_{\mu}U_{\nu}+
q_{\nu}U_{\mu}.
\end{eqnarray}
where $P=\frac{1}{3}(G^1_1+2G^2_2)$ and
\begin{equation}\label{q}
q^\mu=-\frac{e^{-\lambda}}{a(t)^2e^{\frac{\nu}{2}}} H(t)
\nu^{\prime} \delta^{\mu}_r.
\end{equation}
Therefore, in order to have an isotropic solution $G^1_1$ and
$G^2_2$ must meet the $\delta=0$ condition (Eq.~\ref{delta}) which
leads to $P=P_r=P_T=G_1^1=G^2_2$.

A four vector $\xi$ which satisfies
\begin{eqnarray}\label{Ke1}
{\cal L}_\xi g_{\gamma\delta}&=&2\psi g_{\gamma\delta},
\end{eqnarray}
is said to be a conformal killing vector. In the above equation,
$\psi$ is called the conformal factor and for the killing vectors
it takes zero value \cite{KB}. Consider the metric (\ref{csm2}),
and a Killing vector in the form
\begin{eqnarray}\label{KV1}
\xi &=&\alpha(r,t)\frac{\partial}{\partial
t}+\beta(r,t)\frac{\partial}{\partial r}.
\end{eqnarray}
For the conformal killing vector $\xi$ using (\ref{Ke1}), we get
\begin{eqnarray}\label{Ke2}
\psi &=&\frac{\nu'}{2}\beta+\dot{\alpha}\\ \nonumber
\psi &=&\frac{\dot{a}}{a}\alpha+\frac{\lambda'}{2}\beta+\beta' \\
\nonumber \psi &=&\frac{\dot{a}}{a}\alpha+\frac{\mu'}{2}\beta \\
\nonumber 0&=&\alpha'e^{\nu}-a^2e^{\lambda}\dot{\beta}.
\end{eqnarray}
From the second and third equations of the set (\ref{Ke2}) one
reaches
\begin{eqnarray}\label{beta}
\beta(r,t)=f(t)e^{\frac{\mu-\lambda}{2}},
\end{eqnarray}
Therefore, for a given metric, one can get $\xi$ by considering
the conformal factor ($\psi$) and vice versa. In the following
subsections we derive three new classes of solutions.
\subsection{Killing vectors Solutions $\psi=0$\label{sub1}}
Inserting equation (\ref{beta}) into the second and the first
equations of (\ref{Ke2}) we find:
\begin{eqnarray}\label{alpha1}
\alpha &=&-f(t)\frac{a\mu'}{2\dot{a}}e^{\frac{(\mu-\lambda)}{2}} \\
\nonumber
\dot{\alpha}&=&-f(t)\frac{\nu'}{2}e^{\frac{(\mu-\lambda)}{2}}
\end{eqnarray}
respectively. We differentiate with respect to $t$ from the first
equation of (\ref{alpha1}) and comparing the result with the
second equation of (\ref{alpha1}). We get
\begin{eqnarray}\label{alpha2}
f(t)&=&c_1\dot{a} \\ \nonumber \mu &=&\nu+c
\end{eqnarray}
and
\begin{eqnarray}\label{alpha3}
f(t)(2-q)&=&\frac{\dot{f}}{H} \\ \nonumber -\nu+c &=&\mu,
\end{eqnarray}
where $q\equiv-\frac{a\ddot{a}}{\dot{a}^2}$, $c_1$ and $c$ are
arbitrary constants. Consider conditions (\ref{alpha2}), after
taking the derivative of $\alpha$ with respect to $r$ from the
first equation of (\ref{alpha1}) and comparing the result with the
fourth equation of (\ref{Ke2}), we find
\begin{eqnarray}\label{alpha22}
e^{\lambda-\nu}=\mp
\frac{1}{2}(\nu''+\frac{\nu'}{2}(\nu'-\lambda'))
\end{eqnarray}
and
\begin{eqnarray}\label{f}
df=\pm\frac{c_1 dt}{a(t)}.
\end{eqnarray}
Using equation (\ref{ct}), we find:
\begin{eqnarray}\label{alpha22t}
df=\pm c_1 d\eta\rightarrow f(t)=\pm c_1 \eta(t)+c_2.
\end{eqnarray}
Substituting the first set of (\ref{alpha2}) into
(\ref{alpha22t}), one gets;
\begin{eqnarray}\label{alphat}
a(t)=\pm \int \eta(t) dt +\frac{c_2}{c_1}t+c_3.
\end{eqnarray}
Let us substitute $f$ from (\ref{alpha2}) into (\ref{f}) to obtain
\begin{eqnarray}\label{ff}
a\ddot{a}=\pm1,
\end{eqnarray}
which yields
\begin{eqnarray}\label{ff1}
\frac{1}{2}\dot{a}^2=\pm \ln a +C.
\end{eqnarray}
In conclusion we find that, $\xi$ is a killing vector of metric
(\ref{csm2}), when $\mu=\nu+c$, $a(t)$ obeys (\ref{ff}) and the
relation between $\lambda$ and $\nu$ comes from (\ref{alpha22}).
Redshift diverges at $r_0$ if $e^{\nu(r_0)}\rightarrow0$.
Hypersurface located at $r=r_0$ can be timelike, null or even
spacelike. It depends on the value of
$\frac{e^{-\lambda(r_0)}}{a^2(t)}$. In this radius, all of the
curvature scalars diverge and surface area is:
\begin{eqnarray}\label{sa1}
A=\int \sqrt{e^{2\nu(r_0)}e^{2c}a(t)^4sin^2(\theta)}d\theta
d\phi=0.
\end{eqnarray}
Therefore, this should be a naked singularity. Now we consider
(\ref{alpha3}) and following the above recipe to obtain
\begin{eqnarray}\label{fc}
e^{\lambda-\nu}&=&\mp
\frac{1}{2}(-\nu''+\frac{\nu'}{2}(\nu'+\lambda'))\\ \nonumber
a\ddot{a}+2\dot{a}^2&=&\pm1,
\end{eqnarray}
where the second equation yields
\begin{eqnarray}\label{fc1}
\pm1-\frac{1}{Ca^4}=2\dot{a}^2.
\end{eqnarray}
One can write (\ref{fc1}) in the form of
\begin{eqnarray}\label{fc11}
\frac{da}{\sqrt{\pm\frac{1}{2}-\frac{1}{2Ca^4}}}=\pm dt,
\end{eqnarray}
which leads to
\begin{eqnarray}
\sqrt{1\mp a^2}\frac{F(x\sqrt{-\sqrt{\pm
1}},I)-E(x\sqrt{-\sqrt{\pm 1}},I)} {\sqrt{\frac{\pm
a^4-1}{2a^4}}a^2\sqrt{\mp 1}}=\pm t.
\end{eqnarray}
In the above equation, $F(x\sqrt{-\sqrt{\pm 1}},I)$ and
$E(x\sqrt{-\sqrt{\pm 1}},I)$ are incomplete elliptic integrals of
the first and the second kind, respectively. Eventually, $\xi$ is
a killing vector of metric (\ref{csm2}), when $\mu=-\nu+c$,
$\lambda$ and $\nu$ obey the first equation of (\ref{fc}) and
$a(t)$ meets (\ref{fc1}). Similar to the previous case, redshift
(\ref{rs}) diverges at $r_0$ when $e^{\nu(r_0)}\rightarrow 0$.
Also, among curvature scalars, divergence of the Weyl square is
not clear. It depends on the behavior of $e^{\lambda}$, $\nu''$,
$\lambda'$ and $\nu'$ at this radius. The other curvature scalars
will diverge at this radius. For the surface area we get
\begin{eqnarray}\label{s2}
A=\int \sqrt{e^{-2\nu(r_0)}e^{2c}a(t)^4sin^2(\theta)}d\theta d\phi
\rightarrow\infty.
\end{eqnarray}
As the previous case, hypersurface which is located at $r=r_0$ can
be timelike, null or even spacelike and it depends on the value of
$\frac{e^{-\lambda(r_0)}}{a^2(t)}$. Therefore, it is a surface
singularity. Since the redshift singularity either points to the
naked or surface singularities, we think that the redshift
singularity in the killing vector solutions $(\psi=0)$ does not
point to BHs. Briefly, we saw that the naked and surface
singularities cannot live in a universe with arbitrary $a(t)$.
\subsection{CKV solutions $(f(t)=c\neq0)$\label{sub2}}
In this case, from the fourth equation of (\ref{Ke2}) and equation
(\ref{beta}), we get $\alpha'=0$ and $\dot{\beta}=0$,
respectively. Using these results and (\ref{Ke2}), we find
\begin{eqnarray}\label{cm1}
\mu(r) &=&\nu(r)+c\\ \nonumber \alpha(t) &=&a(t)\\
\nonumber \xi^{\alpha}&=&(a(t),ce^{\frac{\mu-\lambda}{2}},0,0),
\end{eqnarray}
and for the conformal factor $\psi(r,t)$, we have
\begin{eqnarray}\label{cf}
\psi(r,t)=\frac{c\nu'}{2}e^{\frac{\mu-\lambda}{2}}+\dot{a}(t).
\end{eqnarray}
Redshift considerations are similar to the case (\ref{alpha2}).
The only major difference is due to the forms of $a(t)$ which are
arbitrary in this case, unlike (\ref{alpha2}) which must meet the
special limitations (\ref{ff}), and should be evaluated from
cosmological considerations. Briefly, this class of solutions
doesn't contain BH.
\subsection{Solutions with $f(t)=0$\label{sub3}}
Using~(\ref{Ke2}) and~(\ref{beta}) we get
\begin{eqnarray}\label{cm}
\beta &=&0\\ \nonumber \alpha &=&a(t)\\ \nonumber
\xi^{\alpha}&=&(a(t),0,0,0)\\ \nonumber \psi &=&\dot{a}(t).
\end{eqnarray}
Therefore irrespective of $\nu$, $\lambda$ and $\mu$, there is a
conformal killing vector $\xi^{\alpha}=(a(t),0,0,0)$ and a
conformal factor $\psi(t)=\partial_{\gamma}\xi^{\gamma}$. $a(t)$
is an arbitrary function of time and must be evaluated from
another parts of physics. We should note that for this class of
solutions, from Eq.~(\ref{Ke2}), it is apparent that the
functional time dependence of $(a(t))$ does not affect the
functional radial dependence of $\nu$, $\lambda$ and $\mu$.
Therefor, when $\beta$ meets the $\beta=0$ condition
Eq.~(\ref{cm}) will be valid for every $a(t)$ independent of
$\nu$, $\lambda$ and $\mu$. For example, one can take it the same
as the scale factor $(a(t))$ of the FRW universe.
\section{Conformally Schwarzschild-de Sitter spacetime\label{sec3}}
In this section, we consider solutions respecting Eq.~(\ref{cm})
and do our calculations in the FRW background. According to the
standard model of cosmology, depending on the equation of state
parameter $\omega=\frac{P}{\rho}$, the scale factor either
increases as a power law $a(t)=At^{\frac{2}{3(\omega+1)}}$ for
$\omega>-1$ or $a(t)=A(t_{br}-t)^{\frac{2}{3(\omega+1)}}$ for
$\omega<-1$, where $t_{br}$ is the Big Rip singularity time and
will happen if the universe is in the phantom regime. For the dark
energy era $(\omega=-1)$, the scale factor is $a(t)=A\ exp\ Ht$.
In the phantom regime, the expansion of the universe ends
catastrophically and everything will ultimately decompose into its
elementary constituents \cite{Mukh}. Simple calculations show that
(\ref{ff1}) and (\ref{fc1}) are not satisfied by the scale factor
of the FRW universe. We take
\begin{eqnarray}\label{fe}
\mu(r)&=&2\ln r \\ \nonumber \nu(r)&=&-\lambda(r)=\ln(1-2m(r)/r).
\end{eqnarray}
We are looking for isotropic solutions of this spacetime. It means
that we want to know the form of $m(r)$ from the isotropy
condition $\delta=0$. Since the anisotropic function ($\delta$) is
independent of time and thus the solutions of the $\delta=0$
equation, the functional time dependence of $(a(t))$ does not
affect the solutions of $\delta=0$. This yields $m(r)=A+Br^3$. It
is apparent that the FRW spacetime is achievable by substituting
$A=B=0$. Consider $B=0$ and $A>0$, then we confront the
Schwarzschild BH embedded in an accelerating universe, which has
been studied by many authors in the literatures for various
accelerating regimes \cite{SD,SD0,SD01,SD',R'}. $A=0$ and $B>0$
yields the de Sitter (dS) spacetime in the static limit $(a(t)
\sim c)$. For the dS spacetime unlike the weak energy condition,
the strong energy condition is violated \cite{P}. Similar to the
dS spacetime, our metric will change its signature at
$r_0=\frac{1}{\sqrt{2B}}$. Also, the divergence of the metric will
happen at this radii and for the surface area at this radius we
have
\begin{eqnarray}
A=\int a(t)^2r_0^2sin(\theta)^2d\theta d\phi=4 \pi R(t)^2r_0^2.
\end{eqnarray}
It is apparent that $\dot{A}\geq0$. Therefore, the second law of
thermodynamics $(\dot{S}\geq0)$ is satisfied \cite{P}. Using
(\ref{nhs}), we see that, just same as the dS spacetime, $r=r_0$
is a null hypersurface and $r>r_0$ and $r<r_0$ point to the
spacelike and the timelike hypersurfaces respectively. Unlike the
Weyl square, the Kretschmann invariant and the Ricci square
diverge at this radius as well as the Ricci scalar. Indeed, the
Weyl tensor is zero for this spacetime showing that this solution
is a conformally flat spacetime \cite{wald}. It is due to this
fact that solutions with $A=0$ are conformal to the dS metric
which is a conformally flat spacetime \cite{P}.
Finally, since this metric is a conformal transformation of the
dS spacetime, its causal structure is the same as that of the dS
spacetime, and therefore we think that the co-moving radii $r=r_0$ points to
a cosmological event horizon like what happen in the similar cases
\cite{SD'}. Consider a co-moving observer, the weak energy
condition yields
\begin{eqnarray}\label{WEC}
-G^0_0\geq 0 \Longrightarrow \dot{a}^2(t)\geq -2B (1-2Br^2),
\end{eqnarray}
which is valid when $B\geq0$. Strong energy condition implies
\begin{eqnarray}\label{SEC}
\frac{1}{2}(3T_1^1-T_0^0)&\geq & 0 \\ \nonumber  \Longrightarrow
2a(t)\ddot{a}(t)+\dot{a}^2(t) & \leq & -4B(1-2Br^2).
\end{eqnarray}
By combining (\ref{WEC}) and (\ref{SEC}), we get
\begin{eqnarray}\label{cws}
2a(t)\ddot{a}(t)-\dot{a}^2(t)\leq0,
\end{eqnarray}
which is a necessary condition for satisfying (\ref{WEC}) and
(\ref{SEC}) simultaneously. This condition is valid when
$\omega\geq-\frac{2}{3}$. So when $\omega\geq-\frac{2}{3}$, strong
and weak energy conditions may be satisfied together. In fact as
sufficient conditions, (\ref{WEC}) and (\ref{SEC}) should be
satisfied separately. It depends on the values of $r$ and $t$ and
can happen when $\omega\geq-\frac{2}{3}$. For the energy flux we
get
\begin{equation}\label{q3}
q^\mu=\frac{2Br\dot{a}(t)}{a(t)^3(1-2Br^2)^{1/2}}\delta^{\mu}_r.
\end{equation}
We see that in the expanding universe $\dot{a}(t)>0$ satisfying
the $B>0$ condition, the energy flux meets the $q^\mu>0$
condition. This fact tells us that everything is ejecting from the
radii $r_0$ during the expansion called the backreaction effect
\cite{Gao2}. The similar result is valid in the dS spacetime
($a(t)=1$) \cite{P}. Since our approach doesn't constrain the
value of $B$, the anti de Sitter (AdS) solution is allowed by our
scheme. In conclusion, solutions with $B\neq0$ and $A=0$, include
the dS and the AdS BHs in a dynamic universe.

Now taking account the condition $A,B\neq0$, we get the
Schwarzschild-de Sitter (SdS) spacetime embedded in the FRW
background, by substitutions $A=m$ and $B=\frac{\Lambda}{6}$. For
the Weyl square we get
\begin{eqnarray}\label{ws}
W=\frac{48A^2}{r^6a(t)^4}.
\end{eqnarray}
Since $A\neq0$ for these solutions, they are not conformal to the
FRW spacetime and therefore, apart of a radial singularity at
$r=0$ which is due to this fact that our spacetime is conformal to
the SdS spacetime, the Weyl square suffers from another
singularity at the bing bang time ($t=0$) produced by the
functional time dependence of the scale factor ($a(t)$). In
addition, Eq.~(\ref{ws}) predicts that this metric is an
asymptotically conformally flat spacetime confirmed by the
asymptotic behavior of the metric and the Weyl tensor in the
$r\gg1$ limit. From Eqs.~(\ref{rs}) and~(\ref{nhs}), we see that
there are two redshift singularities pointing to the null
hypersurfaces and located at
$r^2_c=\frac{1}{2\Lambda}(3(1+\sqrt{1-\frac{4m^2\Lambda}{3}}))$
and
$r^2_e=\frac{1}{2\Lambda}(3(1-\sqrt{1-\frac{4m^2\Lambda}{3}}))$
which are the same as that of the SdS spacetime. Since our
spacetime is conformal to the SdS spacetime, its causal structure
is the same as that of SdS spacetime. The off diagonal elements of
the Einstein tensor $(G_{tr})$ will vanish for large values of
$r$. Therefore, the perfect fluid solution is attainable in this
limit. By evaluating the Ricci scalar we find
\begin{eqnarray}\label{rs1}
R=\frac{R_{FRW}}{1-\frac{2m}{r}-\frac{\Lambda
r^2}{3}}+\frac{4\Lambda}{a^2(t)}.
\end{eqnarray}
From the Einstein equations, the density and the pressure in this
model are
\begin{eqnarray}\label{dp}
\rho(r,t)&=&-T^0_0=\frac{\rho_{FRW}}{1-\frac{2m}{r}-\frac{\Lambda r^2}{3}}+\frac{\Lambda}{a^2(t)}\\
\nonumber
P(r,t)&=&T^i_i=\frac{P_{FRW}}{1-\frac{2m}{r}-\frac{\Lambda
r^2}{3}}-\frac{\Lambda}{a^2(t)},
\end{eqnarray}
where $\rho_{FRW}$ and $P_{FRW}$ are the density and the pressure
of the FRW universe, respectively. By considering
$u^\mu=|-g^{00}|^{1/2}\delta^\mu_t$ and using (\ref{q}), we get
\begin{equation}\label{q2}
q^\mu=-\frac{2(m-\frac{\Lambda}{3}r^3)\dot{a}(t)}{ r^2
a(t)^3(1-\frac{2m(r)}{r})^{1/2}}\delta^{\mu}_r,
\end{equation}
as the radial energy flux which is induced by the background
fluid. Positive (negative) values of $q^\mu$ lead to a mass
decrease (increase) for the BH and at the radius $r$, which
dependents on the values of $m$ and $\Lambda$ \cite{Gao2}. In
fact, $q^\mu<0$ is satisfied when the condition
$r<(\frac{m}{m_0})^{1/3}$ in which $m_0=\frac{\Lambda}{3}$ holds.
The surface area at the radii $r_i\epsilon\{r_e,r_c\}$ is given by
\begin{eqnarray}\label{sa2}
A=\int a(t)^2r_i^2sin(\theta)^2d\theta d\phi=4 \pi a(t)^2r_i^2,
\end{eqnarray}
which increases by the expansion. Also, it is apparent that the
second law of the thermodynamics ($\dot{S}\geq0$) is satisfied.
Based on the properties of the radii $r_i\epsilon\{r_e,r_c\}$ and
since the changes in the metric signature at the radii $r_i$ are
the same as those of the SdS spacetime, we think that there is an event
horizon at co-moving radius $r_e$ where its physical radius is
$\tilde{r}_e=a(t)r_e$ and a cosmological horizon with co-moving
radii $r_c$ and the physical radii $\tilde{r}_c=a(t)r_c$. This
conclusion is supported by this fact that our solution is
conformal to the SdS spacetime \cite{SD',wald}. If we define the
function $f(r)=1-2\frac{m}{r}-\frac{\Lambda}{3}r^2$ and using
(\ref{ah2}), we have $\frac{r_H}{f(r_H)}=\pm\frac{1}{\dot{a}(t)}$
for the apparent horizon radius, which is a fourth order equation
of $r$ and its solutions depend on the values of $m$ and $\Lambda$
and are not straightforward when for the physical radius we get
$\tilde{r}_H=a(t)r_H$. By the slow expansion approximation
$(a(t)\sim C)$, we obtain
\begin{eqnarray}
ds^2\approx -f(\rho)dt^2+\frac{dr^2}{f(\rho)}+\rho^2 d\Omega^2,
\end{eqnarray}
where $\rho\equiv Cr$ and $f(\rho) \equiv
1-\frac{2mC}{\rho}-\frac{\Lambda' \rho^2}{3}$. $\Lambda'$ in the
definition of $f(\rho)$ is $\frac{\Lambda}{C^2}$. By following
\cite{R'}, temperature on the event and cosmological horizons can
be calculated by:
\begin{eqnarray}\label{T}
T_i \simeq \frac{\frac{\partial f}{\partial \rho}|_{r_i}}{4\pi},
\end{eqnarray}
which is compatible with the previous studies \cite{ST,F,R',mr}.
Temperature on the apparent horizon can also be evaluated by
(\ref{SG}). In the limit of zero cosmological constant
$(\Lambda\rightarrow0)$, the results of previous studies are
reproduced \cite{SD,SD0,SD01,R',SD',ST,F}. In fact, the slow
expansion approximation helps us to get an intuitive
interpretation of the BH in the expanding background \cite{R',mr}.

As an another example, we consider a special subclass of solutions
which has the Ricci scalar conformal to the FRW one. For this
class, the condition $R_1=0$ is valid. By choosing (\ref{fe}) and
inserting into $R_1=0$, we get $m(r)=A+\frac{B}{r}$ as general
solution. $A=0$ in the static limit $(a(t)=C)$, points to the
charged massless BHs. Although this solution looks un-interesting,
but it is allowed in the framework of the Yang-Mills theory
\cite{YM}. $B=0$ and $A,B\neq0$ are nothing but the conformally
Schwarzschild and conformally Reissner-Nordst\"{o}m BHs in the FRW
background, respectively \cite{rei,nor}. Physical and
thermodynamical properties of the general solution
$m(r)=A+\frac{B}{r}$, can be found in \cite{R',SD'}.
\section{Concluding remarks\label{sec6}}
In order to find BHs in a dynamic background, we started by the
general form of the static spherically symmetric metrics which
merges smoothly to the dynamic background by a scale factor
$a(t)^2$ and respects certain symmetries. Since we have used the
conformal transformation, the causal structure of the transformed
metric~(\ref{csm2}) is the same as those of the primary
metric~(\ref{ssm}). We found out that some solutions with the
naked and surface singularities cannot be embedded in an arbitrary
dynamical background. In continue, we could find and classify some
special solutions which include various kinds of BHs, within the
same class with the conformal killing vector
$\xi^{\alpha}=a(t)\partial^{t} \delta^{\alpha}_t$ and the
conformal factor $\psi(t)=\dot{a}(t)$, where $a(t)$ is an
arbitrary function of time. For this class of solutions, since
$a(t)$ does not affect the functional radial dependence of the
metric ($\nu$, $\lambda$ and $\mu$), the general properties of the
metric, such as the validity of Eq.~(\ref{q2}) are independent of
the functional time dependence of $a(t)$. We should note that the
quantitative behavior of the spacetime properties and their rates
of changes, such as the energy flux, depend on $a(t)$. In
continue, without loss of generality, we took $a(t)$ the same that
of the FRW spacetime. Among these solutions, the conformally
Schwarzschild BH has special properties. This solution points to
the isotropic fluid and has the Ricci scalar conformal to the
FRW's. The temperature on the redshift singularity surfaces, that
act like horizons, and the apparent horizon have been addressed.
Although the definition of a BH in an expanding universe is vague
\cite{saida}, but our analysis can help us to clarify this
subject. In the early universe, the slow expansion approximation
obviously breaks down and a non-equilibrium analysis will be
needed. Astrophysical motivations for $a(t)$ was not our aim in
this paper. This title could be interesting problem for future
works.
%%%%%%%%%%%%%%%%%%%%%%%%%%%%%%%%%%%%%%%%%%%%%%%%%%%%%%%%%%%%%%%%%%%%%%%
\section*{Acknowledgements}
N. Riazi thanks SBU Research Council for financial support. The
work of H. M. has been supported financially by Research Institute
for Astronomy \& Astrophysics of Maragha (RIAAM) under research
project No.$1/3720-81$.
%%%%%%%%%%%%%%%%%%%%%%%%%%

\end{document}